\documentclass[a4paper]{article}

\usepackage[a4paper,top=3cm,bottom=2cm,left=3cm,right=3cm,marginparwidth=1.75cm]{geometry}

\usepackage{amsmath}
\usepackage{amsthm}
\usepackage{amsfonts}
\usepackage{mathtools} 
\usepackage{bbm} 
\usepackage{bm} 
\usepackage{graphicx}
\usepackage[ruled,vlined,linesnumbered]{algorithm2e}

\newtheorem{theorem}{Theorem}

\def\ddefloop#1{\ifx\ddefloop#1\else\ddef{#1}\expandafter\ddefloop\fi}
\def\ddef#1{\expandafter\def\csname bb#1\endcsname{\ensuremath{\mathbb{#1}}}}
\ddefloop ABCDEFGHIJKLMNOPQRSTUVWXYZ\ddefloop

\def\ddef#1{\expandafter\def\csname c#1\endcsname{\ensuremath{\mathcal{#1}}}}
\ddefloop ABCDEFGHIJKLMNOPQRSTUVWXYZ\ddefloop

\def\ddef#1{\expandafter\def\csname v#1\endcsname{\ensuremath{\boldsymbol{#1}}}}
\ddefloop ABCDEFGHIJKLMNOPQRSTUVWXYZabcdefghijklmnopqrstuvwxyz\ddefloop

\def\ddef#1{\expandafter\def\csname v#1\endcsname{\ensuremath{\boldsymbol{\csname #1\endcsname}}}}
\ddefloop {alpha}{beta}{gamma}{delta}{epsilon}{varepsilon}{zeta}{eta}{theta}{vartheta}{iota}{kappa}{lambda}{mu}{nu}{xi}{pi}{varpi}{rho}{varrho}{sigma}{varsigma}{tau}{upsilon}{phi}{varphi}{chi}{psi}{omega}{Gamma}{Delta}{Theta}{Lambda}{Xi}{Pi}{Sigma}{varSigma}{Upsilon}{Phi}{Psi}{Omega}{ell}\ddefloop

\usepackage[shortlabels]{enumitem} 

\usepackage[backend=bibtex,style=numeric]{biblatex}
\title{To EVM or Not to EVM: \\Blockchain Compatibility and Network Effects}
\author{Ruizhe Jia\\ Columbia University\\ \texttt{rj2536.columbia.edu}
\and 
Steven Yin\thanks{The author was partially supported by a research grant from the Algorand Foundation.}\\Scriptus \\\texttt{steven@scriptus.app}}
\date{July 2022}
\newcommand{\poptAsame}{p^S_{At}}
\newcommand{\poptBsame}{p^S_{Bt}}
\newcommand{\xoptsame}{x^S}
\newcommand{\pioptAsamet}{\pi^S_{At}}
\newcommand{\pioptBsamet}{\pi^S_{Bt}}
\newcommand{\pioptAsame}{\pi^S_{A}}
\newcommand{\pioptBsame}{\pi^S_{B}}

\newcommand{\poptAc}{p^C_{At}}
\newcommand{\poptBc}{p^C_{Bt}}
\newcommand{\xoptc}{x^C}
\newcommand{\pioptAct}{\pi^C_{At}}
\newcommand{\pioptBct}{\pi^C_{Bt}}
\newcommand{\pioptAc}{\pi^C_{A}}
\newcommand{\pioptBc}{\pi^C_{B}}

\newtheorem{Theorem}[theorem]{Theorem}
\newtheorem{Corollary}[theorem]{Corollary}

\newtheorem{Proposition}[theorem]{Proposition}
\newtheorem{Definition}{Definition}
\newtheorem{Assumption}{Assumption}
\begin{document}

\maketitle
\begin{abstract}
We study the competition between blockchains in a \emph{multi-chain} environment, where a dominant EVM-compatible  blockchain (e.g., Ethereum) co-exists with an alternative EVM-compatible (e.g., Avalanche) and an EVM-incompatible  (e.g., Algorand) blockchain. While EVM compatibility allows existing Ethereum users and developers to migrate more easily over to the alternative layer-1, EVM incompatibility might allow the firms to build more loyal and ``sticky'' user base, and in turn a more robust ecosystem. As such, the choice to be EVM-compatible is not merely a technological decision, but also an important strategic decision. In this paper, we develop a game theoretic model to study this competitive dynamic, and find that at equilibrium, new entrants/developers tend to adopt the dominant blockchain. 
To avoid adoption failure, the alternative blockchains have to either (1) directly subsidize the new entrant firms or (2) offer better features, which in practice can take form in lower transaction costs, faster finality, or larger network effects. We find that it is easier for EVM-compatible blockchains to attract users through direct subsidy, while it is more efficient for EVM-incompatible blockchains to attract users through offering better features/products.
\end{abstract}

\section{Introduction}
A commonly held belief in the web3 and blockchain community is that the future is \emph{multi-chain}. In particular, although Ethereum has so far been the dominant blockchain platform in terms of developer and user adoption, there are a number of other layer-1 projects that are gaining significant market shares. Some of them are Ethereum Virtual Machine (EVM) compatible, which makes them relatively easy for both Ethereum developers and users to adopt, since they share the same peripheral tools and applications (e.g. programming languages, wallet software, chrome-plugins, block-explorers). 
Examples of these include Avalanche, NEAR-Aurora, and Polkadot-Moonbeam. There are also blockchains that are not EVM compatible. Examples of such layer-1s include Solana and Algorand. There is often significant friction in users who have already adopted EVM compatible chains adopting non-EVM compatible chains, and vice-versa. There are many sources for such friction. First, EVM incompatibility generally means that both users and developers have to use a different set of tools to access, and develop applications for that blockchain. This means that there is a steep learning curve for users already familiar with the EVM ecosystem to adopt another. Additionally,  many cross-chain solutions (i.e. bridges) launch on EVM-compatible chains first, making it harder for users to move their assets across incompatible chains. As such, for both blockchains and developers, the choice to be EVM-compatible is not just a technological decision, but also an important strategic choice. On the one hand, EVM compatibility makes it easier to attract existing Ethereum users. Large existing user bases also generate positive network effects.  On the other hand, being incompatible with EVM creates a ``lock-in'' effect, which makes users more ``sticky'' in the future, which allows developers to capture more surplus. In this paper, we aim to answer two related questions:

\begin{enumerate}
    \item What types of blockchain platforms will developers and users adopt? Will they choose Ethereum or other EVM-compatible blockchains? or will they adopt EVM-incompatible blockchains?
    \item What strategies can blockchain platforms deploy in order to attract the adoption of developers and users? 
\end{enumerate}
We develop a two stage Bertrand competition game to examine the incentives of firms and users in a multi-chain setting. We show that at equilibrium,  the new entrant firms/developers will adopt the dominant blockchain with the most existing adoption, regardless of whether or not the alternative blockchain is EVM compatible. 

All else being equal, the incentives for a new entrant to build on the dominant blockchain instead of an EVM-compatible alternative chain is intuitive. The dominant blockchain offers the largest existing user base and the largest positive network effect. When the new entrant builds on the same chain as the incumbent, they share the positive network between all of their users. As such, they will not compete as fiercely on price to attract users. 

If 
the new entrant instead develops on the alternative EVM-compatible blockchain, it will no longer share the network effect with the incumbent firm, and will need to engage in intense price competition with the firm on the dominant chain to attract users.  
Since the alternative blockchain has a smaller existing user base and network effect, the firm must be very aggressive in its' pricing strategy in order to attract users, resulting in smaller profits.

What about when the firm builds on an EVM-incompatible blockchain?
Since incompatibility makes it harder for users to switch products, firms might be able to extract more surplus from users if they are built on different incompatible blockchains. Surprisingly, in this case, firms earn even less than the case when they are built on the EVM-compatible alternative chain. The intuition is that even though the firm can now extract more rent from users that it has captured (due to higher user stickiness), it needs to engage in more fierce price competition in the beginning to capture these users in the first place. In fact, the competitive dynamic in the first round of our two stage game is so strong that firms might even offer negative price. This lowers the aggregate payoffs of the firm adopting the EVM-incompatible blockchain. 

To avoid this ``winner takes all" equilibrium and attract the adoption of firms and users, alternative blockchain platforms have to either (1) directly subsidize the new firm, or (2) offer better features (e.g. faster finality, and lower transaction costs).

We show that it is easier for EVM-compatible blockchains to attract users via direct subsidy, while it is more efficient for EVM-incompatible blockchains to attract users by offering better features.
More precisely, compared to EVM-incompatible blockchains, EVM-compatible blockchains can attract the new entrant firm with smaller subsidies. This is because, at equilibrium, when the new firm is built on an EVM-compatible blockchain, the incumbent firm adopts a less aggressive pricing scheme to attract users, which leads to a larger profit margin for both firms.

A less direct way to attract firms to the alternative blockchain is through offering better user experience and better features at the blockchain level.
As mentioned above, this can be in the form of lower transaction fees and better security, but it can also be achieved through more peripheral applications, and other complementary services. When two firms offer a comparable service at the product level, the one that is built on the better blockchain will be able to attract more users. This in turn implies that the firm that is built a superior blockchain will be able to charge a higher price, and thus earn more profits.  As a result, the new entrant firm will choose this alternative blockchain. We show that in our model, EVM-\emph{incompatible} blockchains are able to more efficiently attract more users using this approach, meaning that for each unit of increase in blockchain quality, the EVM-incompatible chain is able to attract more users than a EVM-compatible chain.
\paragraph{Literature review}
The study of network effects \cite{10.2307/1814809} and switching costs was an active research area from the 1980s to the early 2000s, motivated by the then nascent industries of  telecommunication and software development platforms.  A series of papers focused on the effect of switching cost on competition \cite{klemperer1987competitiveness,klemperer1987markets,beggs1992multi}. These papers showed that under different modeling assumptions, the presence of switching cost tend to increase competition between the firms in earlier periods. This is because switching cost increases the firms ability to extract more surplus from their existing adopters, which motivates the firms to compete more fiercely earlier on to capture more new customers.
Building upon these insights, more recent papers \cite{lee2007adoption,zhu2012research} model switching cost as a strategic decision by the platform, and analyze the different settings when firms might want to adopt a lock-in strategy by intentionally creating incompatibility and thus switching costs. For a more comprehensive survey of these topics, see also \cite{belleflamme2018platforms,shy2011short,farrell2007coordination}. Our paper presents a model that combines a few different elements from the existing literature, and presents an analysis of the network effect and switching cost in the context of blockchain applications.

\section{Model Description}
\label{sec: model}
Let there be three blockchain platforms: two EVM-compatible blockchains, $P_1$ and $P_2$, and one EVM-incompatible blockchain, $P_3$. There are two firms, an incumbent $A$ on $P_1$, and a new entrant $B$.

\paragraph{Blockchain Platforms.} Blockchains $P_1$ ,$P_2$, and $P_3$ have existing user bases $N_1$ ,$N_2$, and $N_3$ respectively. A larger user base leads to a larger positive network effect for users, and firms account for this effect when choosing platforms. We assume that $N_1>N_2=N_3$. That is, the EVM-compatible  blockchain $P_1$ has the largest existing user base. Readers can think of $P_1$ as the Ethereum blockchain which has seen the widest adoption amongst users and developers. $P_2$ and $P_3$ are two newer, alternative layer-1 blockchains that compete with Ethereum. $P_2$ is EVM-compatible (e.g. Binance Smart Chain), and $P_3$ is not (e.g. Algorand).

\paragraph{Firms.} Firms $A$ and $B$ build similar applications on blockchain, such as a decentralized exchange (DEX) or a lending service. We assume that the product offered by the firms are inter-operable, meaning that if they are built on the same blockchain, they both benefit from the network effect of each other's users. An example of this would be DEXes on the same chain: with the rise of DEX aggregators, the liquidity found on multiple DEXes on the same blockchain can often be atomically accessed by users at the same time. This would not be true if they were built on two different blockchains. We assume that firm $A$ is already built on the largest blockchain $P_1$ at the start of the game. Firm $B$ is a new entrant who will choose a blockchain to launch its application in period $t=0$. The game will then proceed as a two-stage Bertrand competition, where in each of the two periods $t=1,2$,  the two firms set prices for their services, denoted as $p_{At}, p_{Bt}$ respectively, to maximize their aggregate payoff. We assume that their discount factor is equal to 1 and that the marginal cost of production for both firms is 0. 

\paragraph{Blockchain users.} There is a continuum of users indexed by $x$ on the unit interval $[0,1]$ according to increase personal preference for firm B. All users can gain a benefit of $k>0$ from using the service offered by either firm. Users choose their platform myopically in each period to maximize their payoff. We denote the endogenously determined number of users who choose firms A and B in period $t, t =1,2,$ as $n_{At}, n_{Bt}$ respectively.  The one-period payoff of a user of type $x$ in period $t, t=1,2,$ is given by:

$$U_t(x) = $$
\begin{equation}\label{eq: user payoff}
    \begin{cases}
        \alpha(N_1+n_{At}+n_{Bt})-p_{At}-sx+k, & \text{chooses A, A and B  on $P_1$} \\
        \alpha(N_1+n_{At}+n_{Bt})-p_{Bt}-s(1-x)+k, & \text{chooses B, A and B  on $P_1$} \\
         \alpha(N_1+n_{At})-p_{At}-sx+k, & \text{chooses A,  B  on  } P_i, i\neq 1  \\
        \alpha(N_i+n_{Bt})-p_{Bt}-s(1-x)+k, & \text{chooses B,  B  on } P_i, i\neq 1\\
        0, &\text{chooses neither }.
    \end{cases}
\end{equation}

The parameter $\alpha$ measures the network effect, and the parameter $s>0$ measures the degree of dispersion of consumers' preferences.  As can be seen from users' payoffs, with prices held constant, users benefit the most from the network effect when the two firms are built on the same blockchain. 

We impose the following technical assumption to avoid several trivial corner cases, including the case where network effects are so strong ($\alpha$ is too large) that only one firm would have users in equilibrium, and the case where the benefit of users from using products is so low ($k$ is too low) that most users will drop out of the blockchain platforms. 

\begin{Assumption}
\label{assum: assumption}
The following relationships hold:
\begin{enumerate}
    \item $s>\alpha(2N_1+1)$
    \item $k> 4s+4\alpha(1+N_1+N_2)$
\end{enumerate}
\end{Assumption}

\paragraph{Blockchain incompatibility and user stickiness} 
If a user adopts a product in the first period, then we assume that he faces an additional switching cost if he wishes to use a different product in the second period. We assume that this switching cost is zero if both firms are built on an EVM-compatible blockchain. More specifically, if firm B chooses to build on either $P_1$ or $P_2$, users will have a switching cost of $0$ in the second period. However, if the two firms are built on two incompatible blockchains (i.e. B is built on $P_3$), then users face a non-zero switching cost in the second period. For simplicity, we assume that the switching cost is $+\infty$ in this case. This creates ``stickiness'' amongst the adopters of each firm. Intuitively, firms A and B may take advantage of user stickiness in the second period to extract as much rent as possible. 
\paragraph{Timeline.} The timeline of the model is given as below:
\begin{itemize}
    \item Period 0: Firm B picks a blockchain to offer its product on. 
    \item Period 1:  Firms A and B set prices for their products, and users then make their purchase decision.
    \item Period 2: Firms A and B may adjust the prices for their products, and users then make their purchase decision. 
\end{itemize}

\begin{Definition}[Equilibrium]
    An equilibrium consists of
\begin{enumerate}
    \item Firm B's platform adoption strategy in period 0 and pricing strategy in periods 1 and 2, $d_B = (a, p_{B1}, p_{B2})$, and
    \item firm A's pricing strategy in periods 1 and 2 $d_A =( p_{A1}, p_{A2})$. 
\end{enumerate}
We say that $(d_A,d_B)$ is an equilibrium if it constitutes a subgame perfect Nash equilibrium. 
\end{Definition}

\section{Equilibrium analysis and implications}
\label{sec: analysis}
In this section, we will first focus on the subgame in period 1 and 2, and examine the equilibrium behavior of both the users and the firms under different compatibility settings. Using this, we will derive B's platform adoption decision in period 0. After that, we extend our model to discuss how blockchain platforms can incentivize firms, especially new entrants, to adopt their blockchains



\subsection{Compatible blockchains}
\label{sec: compatible blockchain}

Suppose firm B also adopts an EVM-compatible blockchain, i.e. either $P_1$ or $P_2$. Recall that switching cost is assumed to be zero in this case. As such, users' decisions in period 2 do not depend on their decisions in period 1. From the point of view of firms  A and B, they are playing an identical game of Bertrand competition in each of the two periods. If there exists a unique Nash equilibrium for the one-stage Bertrand competition, denoted as $(p_A,p_B)$, then there exists a unique subgame perfect equilibrium where in each period, firms A and B set the prices as $(p_A,p_B)$.

We will analyze the case when B chooses $P_1$ vs $P_2$ separately, and solve for the equilibrium prices and payoffs in both cases. 

\paragraph{Same blockchain}  Because two firms are built on the same blockchain, the positive network effects for the two firms are identical, so the adoption decision of users are only based on the prices and their individual preferences. Consider the user $\xoptsame$ who is indifferent between the two firms:
$$ \alpha(N_1+n_{At}+n_{Bt})-p_{At}-s \xoptsame+k = \alpha(N_1+n_{At}+n_{Bt})-p_{Bt}-s(1-\xoptsame)+k .$$

Solving the above equation for $\xoptsame$ gives us:
$$\xoptsame = \frac{1}{2}+ \frac{p_{Bt}-p_{At}}{2s}.$$ 
If the marginal user $\xoptsame$ has non-negative surplus, then all users with type $x<x^S$ would choose $A$ and all others will choose $B$. Suppose this is true for now, 
we now consider the Bertrand competition between two firms. 
Firms maximize their one-stage profits:
\begin{align*}
    \pi_{At} &= p_{At} \xoptsame = p_{At} ( \frac{1}{2}+ \frac{p_{Bt}-p_{At}}{2s})\\
    \pi_{Bt} &= p_{Bt} (1-\xoptsame) = p_{Bt} (\frac{1}{2} +  \frac{p_{At}-p_{Bt}}{2s} )
\end{align*}
This will give us a system of first order conditions which we can solve to pin down the equilibrium prices:
\begin{equation}\label{eq: poptsame}
   \poptAsame  = \poptBsame  = s 
\end{equation}
It is easy to verify that all users have non-negative payoff when the prices are set at $(\poptAsame,\poptBsame )$.
Therefore the firms one-period payoffs at equilibrium when they are built on the same blockchain are given by
\begin{equation} \label{eq: payoffsameblockchain one period}
     \pioptAsamet = \pioptBsamet = \frac{s}{2}.
\end{equation}
The total profits of the firms over two periods are then 
\begin{equation} \label{eq: payoffsameblockchain}
      \pioptAsame \coloneqq \pi^S_{A1}+\pi^S_{A2} = \pioptBsame \coloneqq \pi^S_{B1}+\pi^S_{B2} = s. 
\end{equation}

\paragraph{Compatible but different blockchains} Recall that this is when firm $A$ is built on $P_1$ and firm $B$ is built on $P_2$. Unlike the previous case, firms A and B no longer share the same platform, nor do they share the same positive network effect. Instead, they have to maintain the network effects through attracting enough users to their platforms. 

We again first solve for user who is indifferent between the two platforms:
$$ \alpha(N_1+n_{At})-p_{At}-s \xoptc+k = \alpha(N_2+n_{Bt})-p_{Bt}-s(1-\xoptc)+k ,$$ 
Again, if the marginal user $\xoptc$ has non-negative payoff, then everyone with type $x<\xoptc$ would adopt firm A and everyone else would adopt firm $B$.
This means that $n_{At} = \xoptc, n_{Bt} = 1- \xoptc.$
Solving for $\xoptc$ gives us 
\begin{equation}
    \xoptc = \frac{\alpha N_1 - \alpha N_2+(s- \alpha )- p_{At} +p_{Bt}}{2( s-\alpha)}.
    \label{eq: xoptc}
\end{equation}

Note that the expression depends not only on prices but also on the existing user bases. In particular, the larger the existing network effect there is on a firm's blockchain, the more customers that firm is able to capture. Following the same procedure as in the previous case, we can solve for the equilibrium prices of firms as well as their equilibrium payoffs:
\begin{equation}
    \poptAc = s-\alpha + \frac{\alpha(N_1 - N_2)}{3},  \poptBc = s-\alpha + \frac{\alpha(N_2 - N_1)}{3}
\end{equation}

\begin{equation}
    \pioptAct = \frac{(3s-3\alpha + \alpha(N_1 - N_2))^2}{18(s-\alpha)},  \pioptBct = \frac{(3s-3\alpha + \alpha(N_2 - N_1))^2}{18(s-\alpha)}
\end{equation}

It is also easy to verify that all users have positive payoff when the prices are set at $(\poptAc,\poptBc )$.
The total profits over the two periods are simply twice the above amount:
\begin{equation}
    \pioptAc = \frac{(3s-3\alpha + \alpha(N_1 - N_2))^2}{9(s-\alpha)},  \pioptBc = \frac{(3s-3\alpha + \alpha(N_2 - N_1))^2}{9(s-\alpha)}
\end{equation}

The following proposition summarizes the equilibrium results when firms choose compatible blockchains:

\begin{Proposition} \label{prop compatible}
    Suppose firm B chooses an EVM-compatible blockchains (i.e. either $P_1$ or $P_2$). The following holds in equilibrium for the subgame starting in period 1:
    \begin{enumerate}
        \item (Same Chain) If firm B chooses $P_1$ in period 0, then the prices set by firms A and B are $ \poptAsame  = \poptBsame  = s$,for both $t=1,2$. Their aggregate payoffs are $\pioptAsame = \pioptBsame =s.$ Users with $x<\frac{1}{2}$ choose platform A, and the rest choose platform B. 
        \item (Different Chain) If firm B chooses $P_2$ in period 0, then the prices set by firms A and B are $    \poptAc = s-\alpha + \frac{\alpha(N_1 - N_2)}{3},  \poptBc = s-\alpha + \frac{\alpha(N_2 - N_1)}{3}$ for both $t=1,2,$.  Their aggregate payoffs are $ \pioptAc = \frac{(3s-3\alpha + \alpha(N_1 - N_2))^2}{9(s-\alpha)},  \pioptBc = \frac{(3s-3\alpha + \alpha(N_2 - N_1))^2}{9(s-\alpha)}.$ Users with $x<\frac{3s-3\alpha +\alpha(N_1-N_2)}{6(s-\alpha)}$ choose platform A, and the rest choose platform B. 
    \end{enumerate}

\end{Proposition}

\paragraph{Implications of Proposition~\ref{prop compatible}} When firms are built on the same blockchain, users see no difference between the two firms, and the two firms offer the same price and capture the same fraction of the market. When firms are built on two compatible blockchains however, the firm with bigger network effect is at a significant advantage: not only does A capture more of the new market, it is also able to charge more per customer. 
The following corollary is an immediate consequence from comparing the payoff of firms:  

\begin{Corollary} \label{corollary compatible}
The following results hold:
\begin{enumerate}
    \item $\pioptBc < \pioptAc$, i.e., the payoff of firm B is lower than the payoff of firm A when they choose different blockchains. The gap of payoffs $|\pioptBc - \pioptAc|$ increases in $\alpha$.
    \item $\pioptBc < \pioptBsame$, i.e., the payoff of firm B is lower when choosing different blockchain relative to the case when choosing the same blockchain as firm A. Moreover, $\poptBc < \poptBsame$. 
      \item If the difference between existing user bases of two blockchains is small, i.e., $N_1- N_2 \leq 3 $, then $\pioptAc < \pioptAsame, \poptAc < \poptAsame$, that is, the payoff of firm A and the prices set by firm A are lower when firms choose different blockchain, relative to the case when both firms are on the same blockchain. 
\end{enumerate}    
\end{Corollary}

\paragraph{Implications of Corollary \ref{corollary compatible}.} (1) The first result is intuitive. To compensate for its' smaller existing user base, firm B needs to set a lower price, relative to firm A,  to attract users. 
As a result, its profit is also lower than that of firm A in equilibrium. In particular, the stronger the network effect (i.e., $\alpha$), the larger the price gap and profit gap are. (2) Firm B's profit is lower when the B builds on $P_2$ instead of $P_1$.
The reasons are two-folds. 
First, since $N_2<N_1$, firm B starts with a smaller existing user base relative to firm A. Second, by building on a different chain, firm B no longer shares the network effects from the users of firm A. 
This means that firm B has to deploy a more aggressive pricing strategy (which means lower price) to attract users. (3) When the gap between existing user bases are not too big, relative to the case where firms are on the same blockchain, firm A and B need to compete more aggressively when they build on different blockchains. This erodes the profit of firm A because A now needs to charge a low price.

\subsection{Incompatible blockchains}
\label{sec: different incompatible platforms}
We now switch our attention to the case when firm B offers its product on the EVM-incompatible blockchain $P_3$. 
In this case, both firms will have high user stickiness in period 2 due to high switching costs between incompatible blockchains. Intuitively, this will allows the firms to avoid price competition and extract more surplus from users. 

To analyze this formally, we first derive the equilibrium behavior of firms and users in period 2. Since we assume that switching cost is $+\infty$ when chains are incompatible, users cannot switch firms. 
Suppose user $x$ chooses firm A in the first period, his best strategy in the second period is to stay with firm A if the price $p_{A2}$ is low enough for him have non-negative payoff: $\alpha(N_1+n_{A2})-p_{A2}-sx+k \geq 0$. Otherwise, he buys nothing.
We can then write the demand function for firm A as follows:
\begin{equation*}
n_{A2} = 
\begin{cases}
    \min \{ \frac{k+\alpha N_1 - p_{A2}}{s-a}, n_{A1}\} , & \text{if } k \geq p_{A2}- \alpha N_1  \\
       0, & \text{otherwise.}
\end{cases}
\end{equation*}

Given $n_{A1}$, firm $A$ wants to maximize $p_{A2} n_{A2} $ in period 2. Recall $k> 4s+4\alpha(1+N_1+N_2)$ from Assumption~\ref{assum: assumption}.  One can check that the equilibrium price and adoption are given by 
\begin{align}
    p_{A2}^* &= k+\alpha N_1 +(\alpha -s) n_{A1} \label{eq: first helper}\\
    n_{A2}^* &= n_{A1}
\end{align}
That is, firm A would offer the maximum price such that all adopters of A from the first period would weakly prefer to stay with firm A than leaving the system. We can use the same steps to show that given $n_{B1}$, the optimal price and adoption for firm B in period 2 are given by
\begin{align}
    p_{B2}^* &= k+\alpha N_2 +(\alpha -s) n_{B1}\\
    n_{B2}^* &= n_{B1} 
\end{align}
Given the equilibrium behavior of the users and the firms in the second period, we can now derive how firms would price their products in the first period.
Since users are myopic, we can use the same procedure as in Section~\ref{sec: compatible blockchain} to derive that, for a given $p_{A1}, p_{B1}$, the market shares for $A$ and $B$ at equilibrium are:   
\begin{align}
    n_{A1} &= x^I \\
    n_{B1}&= 1- x^I\\
    x^I &= \frac{\alpha N_1 - \alpha N_2+(s- \alpha )- p_{A1} +p_{B1}}{2( s-\alpha)}
    \label{eq: last helper}
\end{align} 
where all users with type $x<x^I$ would adopt $A$ and everyone else would adopt $B$. 
Now, when setting the price in period 1, both firms try to maximize their aggregate payoffs:
\begin{align}
    \pi_{A} = p_{A1}n_{A1} + p_{A2}^*n_{A2}^* \label{eq: incomp total profit A}\\
    \pi_{B} = p_{B1}n_{B1} + p_{B2}^*n_{B2}^* \label{eq: incomp total profit B}
\end{align}
Plugging \eqref{eq: first helper}-\eqref{eq: last helper} into the above equations, we can solve the resulting system of first order conditions for $\pi_A, \pi_B$ to derive the equilibrium prices that the firms offer in the first period:
\begin{align}
    p_{A1}^* = \frac{(10(s-\alpha)-5k -\alpha (N_1+4N_2))}{5}\\ p_{B1}^* = \frac{(10(s-\alpha)-5k -\alpha (4N_1+N_2))}{5}
\end{align}
We can verify that all users have positive in period 1 payoff when the prices are set at $(p_{A1}^*,p_{B1}^* )$. 
 The equilibrium total profits of firms are:
\begin{align}
\pi_A^* = \frac{3(5(s-\alpha) +2\alpha(N_1 - N_2))^2}{100(s-\alpha)}\label{eq: pioptAi}\\
\pi_B^* = \frac{3(5(s-\alpha) +2\alpha(N_2 - N_1))^2}{100(s-\alpha)}\label{eq: pioptBi}
\end{align}

The following proposition characterize the equilibrium results when firm B adopts an incompatible blockchain:

 \begin{Proposition} \label{proposition imcompatible}
     Suppose that firm B adopts platform $P_3$ in period 0. In the subgame that starts in period 1, the following holds:
     \begin{enumerate}
         \item In both periods, users with type $x < \frac{5(s-\alpha)+2\alpha(N_1- N_2)}{10(s-\alpha)}$ chooses firm A, and the rest chooses firm B. 
         \item The prices set by firms A and B in period 1 are 
         $$p_{A1}^* = \frac{(10(s-\alpha)-5k -\alpha (N_1+4N_2))}{5},$$
         $$p_{B1}^* = \frac{(10(s-\alpha)-5k -\alpha (4N_1+N_2))}{5}$$ respectively; the prices set by firms A and B in period 2 are 
         $$p_{A2}^* = k+ \frac{\alpha-s}{2}+\frac{4}{5}N_1 +\frac{1}{5}N_2, p_{B2}^* = k+ \frac{\alpha-s}{2}+\frac{4}{5}N_2 +\frac{1}{5}N_1 $$  
         \item  The aggregate payoffs of firms A and B  are 
         $$\pi_A^* = \frac{3(5(s-\alpha) +2\alpha(N_1 - N_2))^2}{100(s-\alpha)},$$
         $$\pi_B^* = \frac{3(5(s-\alpha) +2\alpha(N_2 - N_1))^2}{100(s-\alpha)}$$ respectively. 
     \end{enumerate}
 \end{Proposition}
\paragraph{Implications of Proposition \ref{proposition imcompatible}.}  We  can see that the equilibrium price of firm  $i\in\{A,B\}$ in period 1 is decreasing in users' stand-alone value for the  product, $k$, as well as the existing user base on its blockchain $N_i$. This might seem counter intuitive, as bigger user base correspond to higher price in the compatible case. The relationship here is reversed due to the fact that the higher the $k$ and $N_i$, the more the firms can extract from users in period 2. As a result, firms are willing lower their prices in period 1 to attract users. 
We can also see that when users' willingness to pay is sufficiently large in period 2 ($k, N_i$ are sufficiently large), the firms are even willing to offer negative prices in the first period! This can be realized in the form of subsidized trading fees, and other liquidity mining incentives. Moreover, the expressions of aggregate payoffs do not contain $k$. This is because all the surplus extracted from users in period 2 are competed away in period 1 in the Bertrand competition.

We are now ready to compare the payoffs of firms in the two cases and characterize the full equilibrium. The following results can be inferred immediately  by comparing $\pi_B^*$ with $ \pioptBc, \pioptBsame$ and comparing $\pi_A^*$ with $ \pioptAc$. 

\begin{Theorem}\label{theorem full equilibrium} The following results hold:
    \begin{enumerate}
        \item $\pi_B^* < \pioptBc , \pi_A^* < \pioptAc$, that is, the payoffs of both A and B firms are lower when firm B chooses $P_3$ relative to the case where firm B chooses $P_2$
        \item In equilibrium, firm B chooses blockchain $P_1$ in period 0. 
    \end{enumerate}
\end{Theorem}

\paragraph{Implications of Theorem~\ref{theorem full equilibrium}} (1) The equilibrium payoffs of both firms in the incompatible case are lower compared with their payoffs in the different but compatible blockchains case. When blockchains are incompatible, users switching cost allows both firms to attract rent from users in period 2. 
As a result, the competition between the firms to attract users in period 1 is more intense, and they need to adopt very aggressive pricing strategies, which lead to a lower overall payoff for both firms. (2) Since $\pi_B^S > \pi_B^C > \pi_B^*$, all else being equal, the new entrant firm B would prefer to build on $P_1$.
Intuitively, this is due to the fact that when firms build on the same blockchain, they share the network effect, and thus do not need to compete as fiercely for market share. When firms adopt different but compatible blockchains, they need to compete more fiercely for market shares because there is no shared mutual benefits between the firms. 
When they adopt different and incompatible blockchains, the need to compete for market share is even stronger (in the first period). This is because firms are essentially monopolies over their adopters in the second period, and so they have strong incentive to capture as much of the market in the first period as possible.

\subsection{How to attract firm adoption}
In the previous section, we established that all else being equal (security, scalability, time to finality, etc), the new entrant firm B will be likely to build on the most popular platform $P_1$ on which firm A is also built.   Viewed from the perspective of a blockchain platforms, our results suggest that alternative layer-1's that have not captured as much adopters as Ethereum will need to offer additional incentives for developers to build on their platform. 

There are two natural ways for blockchain platforms $P_2, P_3$ to attract the adoption of new entrants: (1) directly subsidizing the new entrant firm, (2) improving user experience (e.g., providing higher throughput and thus lower transaction fees, and  providing faster settlement time). In this section, we will extend our model and analyze both options.

\paragraph{Directly subsidizing new firms } Subsidizing developers is a common tactic observed empirically: many alternative layer-1's have significant ecosystem funds for supporting developers\footnote{\url{https://blockworks.co/avalanche-launches-200m-fund-to-support-ecosystem-growth/}}\footnote{\url{https://www.prnewswire.com/news-releases/algorand-foundation-launches-300-million-usd-fund-to-support-defi-innovation-301373155.html}}. In contrast, Ethereum blockchain currently provides less direct subsidization to developers. To understand the effect of subsidization more formally, we extend our baseline model to allow blockchain platforms $P_2, P_3$ to subsidize new firms by offering a one time payment of $c$ if the firm builds on their platform. 
\begin{Proposition} The following results hold:
 \begin{enumerate}
     \item If only blockchain platform $P_2$ offers a subsidy, then in equilibrium, firm B chooses $P_2$ when the subsidy is sufficiently large.   $c>c_2^* = \pioptBsame -\pioptBc$
     \item If only blockchain platform $P_3$ offers a subsidy, then in equilibrium, firm B chooses $P_3$ when the subsidy is sufficiently large.   $c>c_3^* = \pioptBsame - \pi_B^*$. Moreover, $c_3^* > c_2^*$.
 \end{enumerate}
\end{Proposition}
Sufficiently large subsidy induces new entrant to adopt $P_2$ or $P_3$. Since $\pi_B^* < \pioptBc$, it is more costly for EVM-incompatible blockchains to attract adoption because of more intense competition. 

If both $P_2$ and $P_3$ can offer subsidies, and $c_1$ and $c_2$ are their subsidy amount respectively, then firm B would want to choose $P_2$ if and only if
\begin{equation}
    \pioptBc+c_2 \geq \max\{ \pioptBsame, \pi_B^* +c_3 \}.
\end{equation}
On the other hand, $B$ would want to choose $P_3$ if and only if
\begin{equation}
    \pi_B^* +c_3 \geq \max\{ \pioptBsame, \pioptBc+c_2 \}.
\end{equation}
Since the firm's decision depends on both $c_2$ and $c_3$, $P_2$ and $P_3$ have to strategically set their subsidization policies in order attract the firm. However, we will not be providing a formal equilibrium analysis for the subsidy strategy in this paper.

\paragraph{Improving user experience} Platforms can also attract adoption by improving users' stand-alone value of the product ($k$ in our model). Note that the user's experience with a product is a function of both the product itself, as well as the blockchain that the product is built upon. This means that if a blockchain is fundamentally better, users are  more likely to pick products that are built on the better blockchain. This would increase the payoffs of the firms built on the better blockchain. As mentioned before, such improvement take different forms, including: faster finality time, more secure transactions, and lower gas fees. We extend our model to include an additional factor $d$, which represents the extra utility that the user gets from using platform $P_2$ (or $P_3$). Using the same procedure as in our baseline model, we can arrive at the following characterization of the equilibrium results:

\begin{Proposition} \label{prop user improvement 1}
    Suppose that users earn an additional benefit $d $ from using platform $P_2$. Firm B will choose $P_2$ if $d > d_2^* = 3 (s-1) \left(\sqrt{\frac{s}{s-\alpha}}-1\right)+N_1-N_2 $. In such a case, the equilibrium prices  in period 1 and 2 are
    $$  \poptAc = s-\alpha + \frac{\alpha(N_1 - N_2)-d}{3}, 
        \poptBc = s-\alpha + \frac{\alpha(N_2 - N_1)+d}{3}, t=1,2.$$
    Their aggregate payoffs are 
    $$ \pioptAc = \frac{(3s-3\alpha -d+ \alpha(N_1 - N_2))^2}{9(s-\alpha)},
        \pioptBc = \frac{(3s-3\alpha+d + \alpha(N_2 - N_1))^2}{9(s-\alpha)}.$$
    Users with type $x<x^C =\frac{-d+ 3s-3\alpha +\alpha(N_1-N_2)}{6(s-\alpha)}$ would choose firm A, and the rest choose firm B. 
\end{Proposition}
\paragraph{Implication of Proposition \ref{prop user improvement 1}} When blockchain $P_2$ is significantly better than blockchain $P_1$ ($d$ is sufficiently large) , the new entrant will use blockchain $P_2$. This is because with higher extra benefit $d$, more users will choose platform $P_2$, which makes building on $P_2$ more profitable for the firm. The large the $d$, the higher the price $\poptBc$ firm B can charge to users on platform $P_2$, and the higher the payoff of firm B from building on platform $P_2$. This result implies that EVM-compatible blockchains with much lower transaction costs and faster settlement time may have a chance to out-compete the incumbent Ethereum blockchain. In the following Proposition we will state the analogous result for the non EVM-compatible chain:
\begin{Proposition}\label{prop user improvement 2}Suppose that users earn an additional benefit $d $ from using platform $P_3$. Firm B chooses $P_3$ if $d > d_3^* = (s-\alpha) \left(\sqrt{\frac{s}{s-\alpha}}-\frac{5}{2}\right)+N_1-N_2. $  The equilibrium prices in period 1 are 
        \begin{align*}
&p_{A1}^* = \frac{-4d+ (10(s-\alpha)-5k -\alpha (N_1+4N_2))}{5},  \\&p_{B1}^* = \frac{-d+(10(s-\alpha)-5k -\alpha (4N_1+N_2))}{5}, \end{align*} and the equilibrium price in period 2 are:
\begin{align*}   
&p_{A2}^* = k+ \frac{\alpha-s}{2}+\frac{4}{5}N_1 +\frac{1}{5}N_2 +\frac{d}{5},\\  &p_{B2}^* = k+ \frac{\alpha-s}{2}+\frac{4}{5}N_2 +\frac{1}{5}N_1 +\frac{4d}{5}. 
\end{align*}The equilibrium payoffs of firms A and B are
\begin{align*}
    \pi_A^* &= \frac{3(-2d+ 5(s-\alpha) +2\alpha(N_1 - N_2))^2}{100(s-\alpha)},\\
    \pi_B^* &= \frac{3(2d+ 5(s-\alpha) +2\alpha(N_2 - N_1))^2}{100(s-\alpha)}.
\end{align*}
In both periods, users with type $x < x^I =\frac{5(s-\alpha)+2\alpha(N_1- N_2)-2d}{10(s-\alpha)}$ would choose firm A, and the rest would choose firm B. 
\end{Proposition}
\paragraph{Implication of Proposition \ref{prop user improvement 2}.} EVM-incompatible blockchains with better user experience can also attract new entrant firm. Moreover, one unit increase in $d$  increases users adoption ($1- X^I$) by $\frac{1}{5}$. Comparing this with the case for EVM-compatible blockchain in Proposition \ref{prop user improvement 1}, we see that this number is lower at $\frac{1}{6}$. This suggests that it is comparatively \emph{more} effective for incompatible blockchains to attract users and developers via offering a fundamentally better blockchain. 

\paragraph{Increasing network effects.} Another way that platforms can make themselves more attractive to users/firms is through acquiring bigger user bases and thus larger positive network effect (i.e. $N_2$ or $N_3$). Mathematically, this has the same effect as improving user experience, since increasing $N_2$ or $N_3$ is equivalent to increasing the benefit of users on that blockchain, $k$. Therefore we can draw the same conclusions as above for firms' ability to attract users per unit increase of existing network effect.




\section{Conclusion}
Our analysis shows that, absent of external incentives,  developers tend to ``tip'' to one dominant chain due to the network effect. 
Our analysis also highlights interesting strategic differences between EVM-compatible and incompatible blockchains. Compared to incompatible alternative layer-1s, compatible layer-1s can attract firms from the dominant blockchain with less direct subsidy. On the other hand, for every unit increase in the ``quality'' or the existing user base of blockchain, incompatible chains gain more user adoption. 
One key dynamic that we do not capture in our model is negative network effect (i.e. congestion). We leave it to future work for incorporating negative network effect to better predict  adoption patterns between different blockchains. 

\printbibliography
\end{document}